\RequirePackage{fix-cm}
\documentclass[twocolumn]{svjour3}          
\smartqed  
\usepackage{graphicx}
\usepackage{amsmath}
\begin{document}

\title{Phase space analysis of a FRW cosmology in the Maxwell$-$Cattaneo approach\thanks{This work was funded by Comisi\'on Nacional de Investigaci\'on Cient\'{\i}fica y Tecnol\'ogica (CONICYT) through FONDECYT Grant $11140309$ (Y. L.)}
}


\author{Yoelsy Leyva
}


\institute{Yoelsy Leyva \at
              Departamento de F\'isica, Facultad de Ciencias, Universidad de Tarapac\'a, Casilla 7-D, Arica, Chile\\
             \email{yoelsy.leyva@academicos.uta.cl}   
}

\date{Received: date / Accepted: date}

\maketitle

\begin{abstract}
In this work, we present a phase space analysis of a spatially flat Friedmann $-$Robertson$-$Walker (FRW) model in which the dark matter fluid is modeled as an imperfect fluid having bulk viscosity. The bulk viscosity is governed by the Maxwell$-$Cattaneo approach. The rest of the components of the model: radiation and dark energy are treated as perfect fluids. Imposing a complete cosmological dynamics and taking into account a recent constraint on the dark matter equation of state (EOS), we obtain bound on the bulk viscosity. The results point towards the possibility of describing not only the current speed up of the Universe but also the previous matter and radiation dominated eras. 

\end{abstract}

\section{Introduction}
\label{intro}
Nowdays, the cosmological constant ($\Lambda$), among several candidates \cite{Copeland2006,Nojiri2006f,Clifton2012,Bamba2012,Amendola2015,Koyama2016}, remains as the simplest explanation to describe the present period of accelerated expansion of the Universe \cite{Adeothers2016}. However, this description is no free of problems. On the one hand, we have the well known problems of $\Lambda$, which remains unsolved\cite{Weinberg1989a,Padilla2015}\footnote{Some solutions to the cosmological constant problems have been proposed, see for instance \cite{Padmanabhan2013,Nojiri2016}}, and on the other the impossibility of explaining those results that point to a phantom Universe\cite{Adeothers2016}.

Another interesting way to recover accelerated solutions consistent with the latest observations, is by considering a more realistic description of the fluids. This task is carry out by considering imperfect fluids having bulk viscosity \cite{Brevik1994,Zimdahl1996,Nojiri2003,Nojiri2005a,Nojiri2006,Cardone2006,Brevik2011,Brevik2012,Mostafapoor2013,Brevik2015,Brevik2015a,Bamba2016,Sasidharan2016,Laciana2017} \footnote{For extend reviews about bulk viscosity in cosmology see \cite{Gron1990,Maartens1995b,Brevik2014,Brevik2017} and references therein}. The presence of this dissipative mechanism is allowed by the commonly accepted spatial isotropic paradigm of the Universe \cite{Adeothers2016}. At background level, in an expanding spatially flat FRW Universe the presence bulk viscosity introduce a modification to the effective pressure of the cosmic fluids, namely:
\begin{equation}\label{peff}
p_{eff}=p_k+\Pi,
\end{equation}
where $p_k$ is the kinetic pressure of the fluid, and $\Pi$ is bulk viscous pressure. Hence, the background dynamics of the model is modified  because the changes introduced in the density evolution of the viscous fluid. The simplest approach to treat this extra bulk viscosity pressure term in (\ref{peff}) is the Eckart formalism\cite{Eckart1940a}, where, the bulk viscosity pressure is define as:
\begin{equation}\label{peff}
\Pi=-3\zeta H,
\end{equation}
where $H$ is the Hubble parameter and $\zeta$ is the bulk viscosity coefficient. Although this formalism has been widely used at background and perturbative levels \cite{Li2009,Avelino2009,Velten2011,VeltenSchwarz2012,VeltenSchwarzFabrisEtAl2013,Velten2013,AvelinoLeyvaUrena-Lopez2013,CruzLepeLeyvaEtAl2014,Leyva2017}, its main drawback is related with its non-causal behavior allowing superlumincal propagation of the dissipative signal. A causal extension of the Eckart formalism is the so-called Israel-Stewart theory\cite{Israel1979}. The main difference between both formalism is that in the latter the transport equations are differential evolution equations\footnote{See \cite{Maartens1996} for a detailed derivation of the transport equations}, meaning that in the Israel-Stewart case the bulk viscosity pressure obeys:
\begin{equation}\label{IS}
\tau\dot{\Pi}+\Pi=-3\zeta H-\frac{1}{2}\tau\Pi\left(3H+\frac{\dot{\tau}}{\tau}-\frac{\dot{\zeta}}{\zeta}-\frac{\dot{T}}{T}   \right),
\end{equation}
where $\tau$ is the relaxation time and $T$ is the barotropic temperature of the viscous fluid. For barotropic fluids with a constant barotropic index, $\gamma=1+w$, the relaxation time can be reduce to:
\begin{equation}\label{tau}
\tau=\frac{\zeta }{\left(2-\gamma _{v}\right) \gamma_v  \rho _{v}},
\end{equation}
where $\rho_v$ and $\gamma_v$is the energy density and the barotropic index of the viscous fluid. At the same time, the Gibbs integrability condition\cite{Maartens1996} allows to calculate de temperature, $T$, as:
\begin{equation}\label{T}
T\propto\rho_v^{(\gamma_v-1)/\gamma_v}.
\end{equation}
If the near the equilibrium condition $\mid \Pi \mid\ll\rho_v$ is applied to (\ref{IS}) then a truncated version of the Israel$-$Stewart theory is obtained. This causal approach\cite{Zakari1993},  known as Maxwell-Cattaneo equation \cite{Oliveira1988,Pavon1991,Maartens1996}, leads to the following simplified transport equation\footnote{See \cite{Maartens1996,Cruz2017b} for a detailed derivation of the Maxwell$-$Cattaneo equation}:
\begin{equation}
  \dot{\Pi} = -\frac{\Pi \left(3 \gamma _{v} \tau  H+1\right)}{\tau }-\frac{3 \zeta  H}{\tau },
\label{eq:pi}
\end{equation}
Concerning the bulk viscosity coefficient, in the literature, the main ansatz is 
\begin{equation}
\zeta=\zeta(\rho).
\end{equation}
This a direct consequence of considering calculation from the kinetic theory, where transport coefficients are function of powers of the temperature $\zeta=\zeta(T)$\cite{Chapman1991,Hiscock1991,Zakari1993,Jeon1995,Jeon1996,Maartens1996,Bastero-Gil2012}. Another possibility is 
\begin{equation}\label{zeta}
\zeta=\zeta(H),
\end{equation}
where the bulk viscosity coefficient is written as a function of the Hubble paramater\cite{Ren2006,Hu2006,Kremer2012,AvelinoLeyvaUrena-Lopez2013,Mostafapoor2013,Acquaviva2014,Haro2015,Wang2017}, via de Friedmann equations. In the framework of the Eckart theory, this latter ansatz was studied in \cite{AvelinoLeyvaUrena-Lopez2013} to describe the physical viability of a cosmological model in which the dark energy fluid interact with a viscous dark matter. Joint analysis of the observational test and the phase space consistently point to negative values of the bulk viscosity coefficient. These results ruled out any model with this ansatz. In \cite{Acquaviva2014}, the same ansatz was explored in the framework of non$-$linear extension of the Israel$-$Stewart formalism proposed by \cite{Maartens1997}. The authors found that in the case of a viscous radiation is possible to obtain orbits capable of connecting a radiation-dominated era with a matter-dominated transient era to finally evolve to an accelerated expansion solution. More recently, in \cite{Wang2017} the authors proposed three viscous dark energy models in the Eckart formalism to characterize the current speed up of the Universe. The free parameters of the models were constrained with a wide set of cosmological observations, finding a small deviation from the standard cosmological model being able to alleviate the tension in the present value of the Hubble parameter between the Hubble Space Telescope and the global measurement by the Planck Satellite.

The main objective of the present work is to describe the space of solution of an expanding FRW model in the framework of a causal Maxwell-Cattaneo approach. We the aim of generalizing the previous results obtained in \cite{AvelinoLeyvaUrena-Lopez2013}, in the non-causal Eckart formalism, we will choose the bulk viscosity as $\zeta=\zeta(H)$. We will not consider the interaction between the dark matter and dark energy fluids. In this sense, our work will provide a sort of bridge between the non-causal and the causal non$-$linear description studied in \cite{AvelinoLeyvaUrena-Lopez2013} and \cite{Acquaviva2014} respectively. 

This paper is organized as follows: the field equations of the model are presented in Section \ref{model}. In Section \ref{dynamics} we rewrite the field equations by using an appropriate set of dimensionless phase space variables. The corresponding autonomous system is studied by means of the dynamical systems tools. A detailed scheme of the critical points of the system and their existence and stability is shown. We focus our discussion on the viability of a complete cosmological dynamics \cite{AvelinoLeyvaUrena-Lopez2013,LeonLeyvaSocorro2014}. Finally, Section \ref{conclusions} is devoted to conclusions. 

\section{The model}\label{model}
Our starting point is a spatially flat FRW model in which the dark matter fluid is modeled as an imperfect fluid having a bulk viscosity in the framework of Maxwell-Cattaneo approach. Whereas the other fluids considered in the model:  radiation and dark energy, will be treated as perfect fluids. Thus, the field equation can be written as:

  \begin{eqnarray}
    H^2 &=& \frac{8 \pi G}{3}\left( \rho_{\rm r}  + \rho_{\rm dm} +\rho_{\rm de} \right) \, , \label{ConstrainFriedmann} \\
    \dot{\rho}_{\rm dm}  &=& - 3 H \gamma_{dm}\rho_{\rm dm}-3H\Pi   \, , \label{EqCons1} \\
    \dot{\rho}_{\rm r} &=& - 3 H\gamma_{r}\rho_{\rm r} \, , \label{EqCons2} \\
   \dot{\rho}_{\rm de} &=& - 3H\gamma_{\rm de}\rho_{\rm de},  \label{EqConsDE}\\
     \dot{H} &=& -4\pi G \left( \gamma_{r}\rho_{\rm r}  +\gamma_{dm} \rho_{\rm dm} +\gamma_{\rm de} \rho_{\rm de} +\Pi \right)\label{eq:Ra}
  \end{eqnarray}

where $G$ is the Newton gravitational constant, $H$ the Hubble parameter,
$(\rho_{\rm dm}$, $\rho_{\rm r}$, $\rho_{\rm de})$ are the energy
densities of dark matter, radiation and  DE fluid components respectively. Whereas, $\gamma_{\rm de}$ is the barotropic index of the equation of state (EOS)
of DE, which is defined from the relationship $p_{\rm de} = (\gamma_{\rm de} -1)
\rho_{\rm de}$, where  $p_{\rm de}$ is the pressure of DE. 
The term $\Pi$ in (\ref{EqCons1} and \ref{eq:Ra}) is the bulk viscosity pressure term. The evolution of this latter term is given, in the framework of the Maxwell-Cattaneo approach, by (\ref{eq:pi}) identifying  $\rho_v$ as $\rho_{dm}$\footnote{The same identification must be made in (\ref{tau}) for $\gamma_v\rightarrow\gamma_{dm}$.}

As we mentioned before, we take the bulk viscous coefficient $\zeta$ to be proportional to Hubble parameter in the form:
\begin{equation}
  \label{ViscosityDefinition}
 \zeta= \xi_0 H,
\end{equation}
where in order to guarantee nonviolation of the Local Second Law of Thermodynamics (LSLT) \cite{Maartens1996,Zimdahl2000,Misner1973}, $\xi_0\geq0$.

\section{Dynamical System}\label{dynamics}

In order to study all possible cosmological scenarios of the model (\ref{EqCons1}-\ref{eq:Ra}),  we introduce the following dimensionless phase space variables to build an autonomous dynamical system:
\begin{equation}
 x=\Omega_{1}\equiv\frac{8\pi G}{3H^{2}} \rho_{\rm 1}\, ,  y = \Omega_{de}\equiv\frac{8\pi G}{3H^{2}} \rho_{\rm de}, z \equiv\frac{\Pi}{3 H^2};
\label{eq:R}
\end{equation}
then the equation of motion can be written in the following, equivalent, form: 
\begin{eqnarray}
\frac{dx}{dN}&=&3 x y \gamma _{\text{de}}-x (x+4 y-1)+3 (x-1) z \label{as1}\\
\frac{dy}{dN} &=&y \left(3 (y-1) \gamma _{\text{de}}-x-4 y+3 z+4\right)\label{as2}\\
\frac{dz}{dN} &=&3 y z \gamma _{\text{de}}-\frac{3 x z}{\xi _0}-x (z+3)+z(-4 y+3 z+1)\label{as3}
\end{eqnarray}
where the derivatives are with respect to the $e$-folding number $N\equiv \ln a$ and we have reduced one degree of freedom, $\Omega_r=\frac{8\pi G}{3H^2}\rho_r=1-x-y$, by using the Friedmann constraint (\ref{ConstrainFriedmann}).

The phase space of Eqs. (\ref{as1}-\ref{as2}) can be defined as
\begin{eqnarray}
  \Psi &=& \{(x,y): 0 \leq 1-x-y \leq 1 ,  0\leq x \leq1 , \nonumber\\
  && 0 \leq y \leq 1\} \,, \label{eq:space}
\end{eqnarray}
where we have imposed the conditions that fluid components be positive, definite, and bounded at all times. 

In order to discuss the dynamics associated with the critical points of the autonomous system (\ref{as1}-\ref{as3}), we need to introduce some cosmological parameter of interest, such as the deceleration parameter ($q=-(1+\dot{H}/H^2)$) and the effective EOS ($w_{eff}$) in terms of the dimensionless variables (\ref{eq:R}). Following this, they can be expressed as:
\begin{eqnarray}
  q &=&\frac{1}{2} \left(3 y \gamma _{\text{de}}-x-4 y+3 z+2\right)\label{qq}\\
  w_{eff} &=& \frac{1}{3} (-x-4 y+1)+z+y \gamma _{\text{de}}\label{w}
\end{eqnarray}

The autonomous system (\ref{as1}-\ref{as3}) admits five critical points which are shown in Table \ref{tab1}, whereas the corresponding eigenvalues of the linear perturbation matrix and some important parameters are displayed in Table \ref{tab2}. In order to discuss the existence and stability behavior of the critical points $P_1$$-$$P_5$, here we summarized their basic properties.

\subsection{Critical points and stability}\label{subscp}

\begin{table*}
\caption{Location, existence conditions
    according to the physical phase space (\ref{eq:space}), and
    stability of the critical points of the autonomous system
    (\ref{as1})-(\ref{as2}). The
    eigenvalues of the linear perturbation matrix associated to each
    of the following critical points are displayed in
    Table~\ref{tab2}. We have introduced the definition $A=\sqrt{4 \xi _0^2+1}$}
\label{tab1}
\begin{tabular*}{\textwidth}{@{\extracolsep{\fill}}lccccc@{}}
\hline
$P_i$ & $x$ & $y$ & $z$ & Existence & Stability \\
\hline
$P_{1}$ & $1$ & $0$ & $\frac{1+A}{2 \xi _0}$  & Always  &  Unstable if $\gamma _{\text{de}}<\frac{2}{3}\land \xi _0>0$   \\ 

 $P_{2}$ & $0$ & $0$ & $0$  & Always & Saddle  \\ 
	       
$P_{3}$ & $1$ & $0$ & $\frac{1-A}{2 \xi _0}$  & Always  & Stable if $0<\gamma _{\text{de}}<\frac{2}{3}\land \xi _0>\frac{\gamma_{\text{de}}-1}{\gamma _{\text{de}}^2-2 \gamma _{\text{de}}}$ \\ 
                    &        &       &         &               &  Saddle see section \ref{subscp}\\

$P_{4}$ & $0$ & $1$ & $0$  & Always  & Stable if  $\gamma_{de}<0$\\
              &        &       &         &               & Saddle if  $0<\gamma_{de}<\frac{2}{3}$\\
              
$P_{5}$ & $\frac{\xi _0 \left(\gamma _{\text{de}}-2\right) \gamma
   _{\text{de}}}{\gamma _{\text{de}}-1}$ & $1-\frac{\xi _0 \left(\gamma _{\text{de}}-2\right) \gamma
   _{\text{de}}}{\gamma _{\text{de}}-1}$ & $\xi _0 \left(\gamma _{\text{de}}-2\right) \gamma _{\text{de}}$  & See section \ref{subscp}  & Stable if  $0<\gamma _{\text{de}}<\frac{2}{3}\land 0< \xi _0<\frac{\gamma
   _{\text{de}}-1}{\gamma _{\text{de}}^2-2 \gamma _{\text{de}}}$\\
\hline
\end{tabular*}
\end{table*}

\begin{table*}
\caption{Eigenvalues and some basic physical
    parameters for the critical points listed in Table~\ref{tab1}, see
    also Eqs.~(\ref{qq}) and~(\ref{w}). We have introduced the definition $B=\sqrt{\gamma _{\text{de}} \left(\gamma _{\text{de}} \left(-4 \xi _0
   \left(\gamma _{\text{de}}-1\right) \left(\gamma
   _{\text{de}}-2\right){}^2+5 \gamma _{\text{de}}^2-20 \gamma
   _{\text{de}}+28\right)-16\right)+4}$}
\label{tab2}
\begin{tabular*}{\textwidth}{@{\extracolsep{\fill}}lcccccc@{}}
\hline

$P_i$&$\lambda_1$ & $\lambda_2$ & $\lambda_3$ & $\Omega_r$ & $w_{eff}$ & $q$ \\ \hline

  $P_{1}$ &$\frac{3 A}{\xi _0}$ & $\frac{3 \left(A-2 \xi _0 \left(\gamma _{\text{de}}-1\right)+1\right)}{2
   \xi _0}$&$\frac{3 A-2 \xi _0+3}{2 \xi _0}$ & $0$&$\frac{1+A}{2 \xi _0}$ & $\frac{1}{2} \left(1+\frac{3 (A+1)}{2 \xi _0}\right)$\\

$P_{2}$ & $4$ & $-2$ & $4-3\gamma_{de}$ & $1$ & $\frac{1}{3}$ & $1$ \\ 

  $P_{3}$ & $-\frac{3 A}{\xi _0}$ & $\frac{-3 A-6 \xi _0 \left(\gamma _{\text{de}}-1\right)+3}{2 \xi _0}$ &$\frac{-3 A-2 \xi _0+3}{2 \xi _0}$ &$0$ &$\frac{1-A}{2 \xi _0}$ &$\frac{1}{2} \left(1+\frac{3 (1-A)}{2 \xi _0}\right)$ \\ 
  
    $P_{4}$ & $3(-2+\gamma_{de})$ & $3\gamma_{de}$ & $-4+3\gamma_{de}$ & $0$ & $-1+\gamma_{de}$ & $-1+\frac{3\gamma_{de}}{2}$ \\   
    
    $P_{5}$ & $-4+3\gamma_{de}$ & $\frac{3 \left(-B+\left(\gamma _{\text{de}}-2\right) \gamma
   _{\text{de}}+2\right)}{2 \left(\gamma _{\text{de}}-1\right)}$ & $\frac{3 \left(B+\left(\gamma _{\text{de}}-2\right) \gamma
   _{\text{de}}+2\right)}{2 \left(\gamma _{\text{de}}-1\right)}$ & $0$ & $-1+\gamma_{de}$ & $-1+\frac{3\gamma_{de}}{2}$ \\   
\hline
\end{tabular*}
\end{table*}

Critical point $P_1$ represents a solution  dominated by the dark matter component ($x=\Omega_m=1$) and always exist. However, a background level this solution behave as stiff matter, namely $1<w_{eff}<\infty$ ($2<q<\infty$). The limit case of $w_{eff}\approx 1$ corresponds to extremely high values of the bulk viscosity parameter, $\xi_0\rightarrow\infty$, while $w_{eff}\approx \infty$ corresponds to very small values of it, $\xi_0\rightarrow 0$. This solution is always unstable for all realistic dark energy fluid, $w_{de}<-1/3$ ($\gamma_{de}<2/3$).

$P_2$ corresponds to a decelerating solution ($w_{eff}=1/3$, $q=1$) dominated by the radiation component, $\Omega_r=1$ and always exists. As Table \ref{tab2} shows, this point is always saddle since its eigenvalues have opposite signs. 

$P_3$  represents a dark matter solution ($x=\Omega_m=1$) and always exists. But unlike $P_1$, a background level this solutions is able to mimic decelerated and accelerated solutions, namely $-1<w_{eff}\leq 0$. The standard matter-domination period , $w_{eff}=0$ and $q=1/2$, corresponds to the limit $\xi_0=0$ while the quintessence like solutions, $-1<w_{eff}<-1/3$, are recovery for $\xi_0>3/8$. As Tables \ref{tab1} shows, these accelerated solutions are possible in the absence of the dark energy component, $\Omega_{de}=y=0$. Critical point $P_3$ exhibits two different stability behaviors:
\begin{enumerate}
\item Decelerating region, $-1/3<w_{eff}\leq0$:
\begin{itemize}
\item Saddle if $\gamma _{\text{de}}<\frac{2}{3}\land 0<\xi _0<\frac{3}{8}$
\end{itemize}
\item Quintessence region, $-1<w_{eff}<-1/3$:
\begin{itemize}
\item Stable if $0<\gamma _{\text{de}}<\frac{2}{3}\land \xi _0>\frac{\gamma_{\text{de}}-1}{\gamma _{\text{de}}^2-2 \gamma _{\text{de}}}$
\item Saddle if $\left(\gamma _{\text{de}}\leq 0\land \xi _0>\frac{3}{8}\right)\lor\newline\left(0<\gamma _{\text{de}}<\frac{2}{3}\land \frac{3}{8}<\xi_0<\frac{\gamma _{\text{de}}-1}{\gamma _{\text{de}}^2-2 \gamma_{\text{de}}}\right)$
\end{itemize}

\end{enumerate}

Critical point $P_4$ corresponds to a solution dominated by the dark energy fluid $\Omega_{de}=1$ and always exists. For all realistic dark energy fluid, $w_{de}<-1/3$, $P_4$ represents an accelerated solutions. In the phantom region, $w_{de}<-1$($\gamma_{de}<0$) this solution is stable. Whereas, in the quintessence region, $-1<w_{de}<2/3$ ($0<\gamma_{de}<2/3$), $P_4$ behave as saddle solution. 

Finally, $P_5$ represents and scaling solution between dark matter and dark energy and exists when:
\begin{itemize}
\item $\left(\gamma_{\text{de}}=0\land \xi _0\geq 0\right)\lor$
   \item $\left(0<\gamma_{\text{de}}<\frac{2}{3}\land 0\leq \xi _0\leq \frac{\gamma_{\text{de}}-1}{\gamma _{\text{de}}^2-2 \gamma _{\text{de}}}\right)$
\end{itemize}
A background level, this critical point is able to mimic accelerated solutions in the quintessence and de Sitter regions:

\begin{enumerate}
\item Quintessence region, $-1<w_{eff}<-1/3$ ($0<\gamma_{de}<2/3$):
\begin{itemize}
\item Stable if $0\leq \xi _0<\frac{\gamma _{\text{de}}-1}{\gamma _{\text{de}}^2-2 \gamma_{\text{de}}}$. 
\end{itemize}
\item de Sitter region, $w_{de}=-1$ ($\gamma_{de}=0$). As Table \ref{tab1} and \ref{tab2} show, in this limit case,  $P_5$ and $P_4$ are the same critical point. Under this value of the barotropic index this point behaves as a nonhyperbolic critical point with a two dimensional stable manifold ($\lambda_1=-4$, $\lambda_2=0$ and $\lambda_3=-6$). In this case, the standard linear dynamical systems analysis fails to be applied, then we will rely our analysis on numerical inspection of the phase portrait.

\end{enumerate}

\subsection{Cosmology evolution}\label{cosmoe1}

In order to make a complete description of the evolution of the Universe, we must demand that our model is able to reproduce three different periods from early to late times, namely:
\begin{itemize}
\item radiation-dominated era (RDE) $\rightarrow$ $\Omega_r=1$
\item matter-dominated era (MDE), $\rightarrow$ $\Omega_{dm}=1$
\item period of accelerated expansion, $\rightarrow$ $w_{eff} \approx -1$
\end{itemize}
From the dynamical system point of view, every one of this periods is represented by a critical point and the corresponding transitions between them correspond to heteroclinic orbits. 

As Table \ref{tab1} shows, the unstable nature of the dark matter stiff solution $P_1$, guarantees it can be, at early times, the source of any solution in the phase space for any value of the bulk viscosity parameter, $\xi_0$, and for every realistic dark energy fluid ($\gamma_{de}<2/3$). Its stiff matter behavior can be understood because of the effect of the viscous pressure $\Pi$. As Table \ref{tab1} shows, at early times, the viscous pressure contributed with a positive pressure as $A=\sqrt{4\xi_0^2+1}>1\rightarrow z=\frac{\Pi}{3H^2}>0$\footnote{Recall that $z=\frac{1+A}{2\xi_0}$ for $P_1$}. 

The condition for a radiation-dominated era, RDE, ($\Omega_r=1$) is fulfilled by $P_2$. Its saddle behavior, independently of the values of $\xi_0$ and $\gamma_{de}$, means that is possible to find appropriate initial conditions allowing us to connect this RDE critical point with the unstable stiff matter solution $P_1$. In the case of this RDE solution, the viscous pressure has no effect in its existence or stability behavior, namely $z=0$, thus $\Pi=0$.

In order to describe the formation of cosmic structures, the existence of a MDE, at intermediate stage of the evolution of the Universe, is needed. This period of matter domination can be recovered by $P_1$ or $P_3$ . As we mentioned before, the unstable nature of $P_1$ and the requirement of a previous period of radiation domination, ruled out this point as a true candidate to describe the MDE. On the other hand, $P_3$ is able to reproduce a decelerating solution dominated by the dark matter component if $0<\xi _0<3/8$. 
Following  \cite{Leyva2017}, we will constrain the possible values of $\xi_0$ to fulfill a true MDE ($w_{eff}\approx0$) by considering a bound on the dark matter EOS by \cite{ThomasKoppSkordis2016} using the latest Planck results \cite{Adeothers2016}. This result establishes,  a $3\sigma$, that:$-0.000896<w_{dm}<0.00238$. Combining this result with the constraints found in the previous section allowing a saddle behavior for $P_3$, we obtain the region:
\begin{equation}\label{rxi}
0<\xi _0<0.000896,
\end{equation}
meaning that a true MDE demands a very small contribution of the bulk viscosity\footnote{This result is the same found in \cite{Leyva2017} in the Eckart framework. 
}.  Another interesting characteristic of $P_3$ is that, if the conditions $0<\gamma _{\text{de}}<\frac{2}{3}\land \xi _0>\frac{\gamma_{\text{de}}-1}{\gamma _{\text{de}}^2-2 \gamma _{\text{de}}}$ are fulfilled, then $P_3$ is able to reproduce an stable accelerated solution in the quintessence region being a candidate to explain the late time speed up of the Universe. This requieres higher values of the bulk viscosity parameter than those required for a MDE era, hence, the viscous pressure contributed with an appreciable negative pressure ($z<0$, $\Pi<0$). Despite this characteristic, as Table \ref{tab1} and \ref{tab2} show, this possible behavior has to be ruled out because of the impossibility of finding another critical point capable of reproducing a true MDE.

\begin{figure}
 \includegraphics[width=0.48\textwidth]{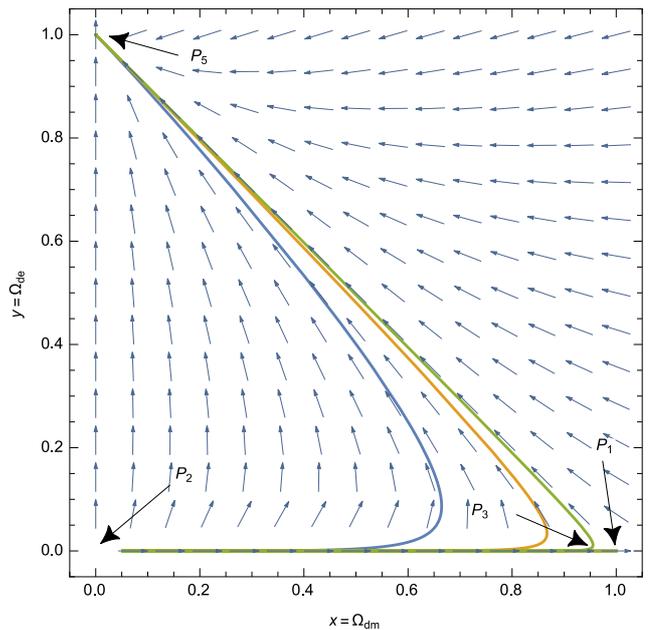}
\caption{Projections of some orbits of the phase space of the autonomous system (\ref{as1}-\ref{as3}) in the subspace (x,y). The free parameter ($\xi_0$, $\gamma_{de}$) have been chosen as ($0.0008$, $0,04$). All the orbits start at the stiff matter solution $P_1$ evolving to a RDE solution $P_2$. From this latter critical point the orbits approach to the saddle MDE solution $P_3$. For the previous choice of the free parameter, the attractor solution is always the quintessence solution $P_5$.} \label{f1}
\end{figure}

In terms of the late time evolution of the Universe, there is two critical points capable of providing accelerated solutions, namely $P_4$ and $P_5$. The first one represents a pure dark energy solution ($\Omega_{de}=1$) with a null contribution of the bulk viscosity pressure, $z=0$. Whereas, the latter corresponds to a scaling solution between dark matter and dark energy where the bulk viscosity contributes with a negative pressure term, namely $z<0$, $\Pi<0$. In both cases, the effective EOS parameter depends on the barotropic index of the dark energy fluid: $w_{eff}=1-\gamma_{de}$. This means that the only possible way to achieve a phantom solution is if the dark energy fluid is phantom, otherwise is impossible. Even in the case of $P_5$, where the bulk viscosity pressure is negative ($z<0$, $\Pi<0$), the effective EOS parameter is independent of the value of $\Pi$.  As Tables \ref{tab1} and \ref{tab2} shows, in the case of a phantom dark energy fluid ($\gamma_{de}<0$), the only possible late time scenario is the stable phantom solution $P_4$\footnote{in this case $P_5$ do not exist in the phase space.}.  However, in the case of a quintessence dark energy fluid ($0<\gamma_{de}<2/3$) the late time behavior of the model allows a transition between two quintessence solutions, from $P_4$ (saddle) to $P_5$(stable). If we impose the condition (\ref{rxi}) to ensure a true MDE and take into account the latest constraint on the value of the dark energy EOS \cite{Adeothers2016}, $P_4$ and $P_5$ become almost indistinguishable in the phase space as $x\rightarrow0$, $y\rightarrow1$ and $z\rightarrow0$. 

Figure \ref{f1} shows projections of some example orbits in the plane ($x=\Omega_{dm}$,
$y=\Omega_{de} $) to illustrate the above scenario.

\section{Concluding remarks}\label{conclusions}

In this work, we studied the dynamics of a model of the Universe filled with radiation, dark matter, and dark energy. The dark matter component was treated as an imperfect fluid having bulk viscosity in the framework of a Maxwell-Cattaneo approach. Whereas the remaining fluids were considered as perfect fluids. The bulk viscosity was taken as proportional to the Hubble parameter $\xi_0\propto H$ \cite{Kremer2012,AvelinoLeyvaUrena-Lopez2013,Acquaviva2014}. 

In order to investigate the asymptotic evolution of the model, we performed a dynamical system analysis. This analysis was based on the requirement of a transition from a RDE to a MDE, at early and intermediate stages, to finally converge to an accelerated solution at late times. This statement, together with a recent estimation of the dark matter EOS reduce the bulk viscosity parameter to a very small region, namely $0<\xi_0<0.000896$. This result reproduces similar results obtained in the non-causal Eckart framework in \cite{VeltenSchwarz2012,VeltenSchwarzFabrisEtAl2013,CruzLepeLeyvaEtAl2014,Leyva2017} and more recently by \cite{Lepe2017} in the Israel$-$Stewart formalism.

In the absence of a dark energy fluid, the model is able to reproduce a stable accelerated solution in the quintessence region due to the contribution of the bulk viscosity pressure. However, this solution is ruled out as a possible answer to the present period of accelerated expansion of the Universe since it is impossible to reproduce previous periods of radiation and dark matter domination. 

From the stability point of view, the late time attractor, compatible with previous stages of radiation and matter dominations, is always an accelerated solution. This can be a phantom($P_4$) or quintessence ($P_5$) solutions
. The nature of both possible solutions depends only on the barotropix index of the dark energy fluid, namely $\gamma_{de}$. Hence, the only way this model is able to cross the phantom divide ($w_{eff}<-1$) is taking a phantom fluid ($\gamma_{de}<0$) as a dark energy source. Thus the possible transitions are: $P_2$(RDE)$\rightarrow$ $P_3$(MDE)$\rightarrow$$P_4$ or $P_2$(RDE)$\rightarrow$ $P_3$(MDE)$\rightarrow$$P_4$$\rightarrow$$P_5$.

In all the cases, the source of any orbits in the phase space is the stiff matter solution $P_1$\footnote{This critical point is equivalent to $R_{2+}$ in \cite{Lepe2017}, where the Israel$-$Stewart formalism is used for a different parametrization of the bulk viscosity.}. Contrary to the late time case, this point exhibit a contribution of a positive bulk viscosity pressure. This explains the extremely high values taken by the $z$ variable, and consequently by the bulk viscosity pressure $\Pi$, at this period of the evolution of the Universe in this model.

%
%


\begin{thebibliography}{10}

\bibitem{Copeland2006}
Edmund~J. Copeland, M.~Sami, and Shinji Tsujikawa.
\newblock {Dynamics of dark energy}.
\newblock {\em Int. J. Mod. Phys.}, D15:1753--1936, 2006.

\bibitem{Nojiri2006f}
Shin'ichi Nojiri and Sergei~D. Odintsov.
\newblock {Introduction to modified gravity and gravitational alternative for
  dark energy}.
\newblock {\em eConf}, C0602061:06, 2006.
\newblock [Int. J. Geom. Meth. Mod. Phys.4,115(2007)].

\bibitem{Clifton2012}
Timothy Clifton, Pedro~G. Ferreira, Antonio Padilla, and Constantinos Skordis.
\newblock {Modified Gravity and Cosmology}.
\newblock {\em Phys. Rept.}, 513:1--189, 2012.

\bibitem{Bamba2012}
Kazuharu Bamba, Salvatore Capozziello, Shin'ichi Nojiri, and Sergei~D.
  Odintsov.
\newblock {Dark energy cosmology: the equivalent description via different
  theoretical models and cosmography tests}.
\newblock {\em Astrophys. Space Sci.}, 342:155--228, 2012.

\bibitem{Amendola2015}
Luca Amendola and Shinji Tsujikawa.
\newblock {\em {Dark Energy}}.
\newblock Cambridge University Press, 2015.

\bibitem{Koyama2016}
Kazuya Koyama.
\newblock {Cosmological Tests of Modified Gravity}.
\newblock {\em Rept. Prog. Phys.}, 79(4):046902, 2016.

\bibitem{Adeothers2016}
P.~A.~R. Ade et~al.
\newblock {Planck 2015 results. XIII. Cosmological parameters}.
\newblock {\em Astron. Astrophys.}, 594:A13, 2016.

\bibitem{Weinberg1989a}
S.~Weinberg.
\newblock The cosmological constant problem.
\newblock {\em Reviews of Modern Physics}, 61:1--23, January 1989.

\bibitem{Padilla2015}
Antonio Padilla.
\newblock {Lectures on the Cosmological Constant Problem}.
\newblock 2015.

\bibitem{Padmanabhan2013}
T.~Padmanabhan and Hamsa Padmanabhan.
\newblock {CosMIn: The Solution to the Cosmological Constant Problem}.
\newblock {\em Int. J. Mod. Phys.}, D22:1342001, 2013.

\bibitem{Nojiri2016}
Shin'ichi Nojiri.
\newblock {Some solutions for one of the cosmological constant problems}.
\newblock {\em Mod. Phys. Lett.}, A31(37):1650213, 2016.

\bibitem{Brevik1994}
Iver~H. Brevik and L.~T. Heen.
\newblock {Remarks on the viscosity concept in the early universe}.
\newblock {\em Astrophys. Space Sci.}, 219:99, 1994.

\bibitem{Zimdahl1996}
Winfried Zimdahl.
\newblock {'Understanding' cosmological bulk viscosity}.
\newblock {\em Mon. Not. Roy. Astron. Soc.}, 280:1239, 1996.

\bibitem{Nojiri2003}
Shin'ichi Nojiri and Sergei~D. Odintsov.
\newblock {Quantum de Sitter cosmology and phantom matter}.
\newblock {\em Phys. Lett.}, B562:147--152, 2003.

\bibitem{Nojiri2005a}
Shin'ichi Nojiri and Sergei~D. Odintsov.
\newblock {Inhomogeneous equation of state of the universe: Phantom era, future
  singularity and crossing the phantom barrier}.
\newblock {\em Phys. Rev.}, D72:023003, 2005.

\bibitem{Nojiri2006}
Shin'ichi Nojiri and Sergei~D. Odintsov.
\newblock {Unifying phantom inflation with late-time acceleration: Scalar
  phantom-non-phantom transition model and generalized holographic dark
  energy}.
\newblock {\em Gen. Rel. Grav.}, 38:1285--1304, 2006.

\bibitem{Cardone2006}
Vincenzo~F. Cardone, C.~Tortora, A.~Troisi, and S.~Capozziello.
\newblock {Beyond the perfect fluid hypothesis for dark energy equation of
  state}.
\newblock {\em Phys. Rev.}, D73:043508, 2006.

\bibitem{Brevik2011}
I.~Brevik, E.~Elizalde, S.~Nojiri, and S.~D. Odintsov.
\newblock {Viscous Little Rip Cosmology}.
\newblock {\em Phys. Rev.}, D84:103508, 2011.

\bibitem{Brevik2012}
I.~Brevik, R.~Myrzakulov, S.~Nojiri, and S.~D. Odintsov.
\newblock {Turbulence and Little Rip Cosmology}.
\newblock {\em Phys. Rev.}, D86:063007, 2012.

\bibitem{Mostafapoor2013}
Nouraddin Mostafapoor and Oyvind Gron.
\newblock {Bianchi Type-I Universe Models with Nonlinear Viscosity}.
\newblock {\em Astrophys. Space Sci.}, 343:423--434, 2013.

\bibitem{Brevik2015}
Iver Brevik.
\newblock {Viscosity-Induced Crossing of the Phantom Barrier}.
\newblock {\em Entropy}, 17:6318--6328, 2015.

\bibitem{Brevik2015a}
I.~Brevik, V.~V. Obukhov, and A.~V. Timoshkin.
\newblock {Dark Energy Coupled with Dark Matter in Viscous Fluid Cosmology}.
\newblock {\em Astrophys. Space Sci.}, 355:399--403, 2015.

\bibitem{Bamba2016}
Kazuharu Bamba and Sergei~D. Odintsov.
\newblock {Inflation in a viscous fluid model}.
\newblock {\em Eur. Phys. J.}, C76(1):18, 2016.

\bibitem{Sasidharan2016}
Athira Sasidharan and Titus~K. Mathew.
\newblock {Phase space analysis of bulk viscous matter dominated universe}.
\newblock {\em JHEP}, 06:138, 2016.

\bibitem{Laciana2017}
Carlos~E. Laciana.
\newblock {A causal viscous cosmology without singularities}.
\newblock {\em Gen. Rel. Grav.}, 49(5):62, 2017.

\bibitem{Gron1990}
O.~Gron.
\newblock {Viscous inflationary universe models}.
\newblock {\em Astrophys. Space Sci.}, 173:191--225, 1990.

\bibitem{Maartens1995b}
Roy Maartens.
\newblock {Dissipative cosmology}.
\newblock {\em Class. Quant. Grav.}, 12:1455--1465, 1995.

\bibitem{Brevik2014}
Iver Brevik and Øyvind Grøn.
\newblock {Relativistic Viscous Universe Models}.
\newblock In Anderson Travena and Brady Soren, editors, {\em Recent Advances in
  Cosmology}, pages 97--127. 2014.

\bibitem{Brevik2017}
Iver Brevik, Øyvind Grøn, Jaume de~Haro, Sergei~D. Odintsov, and Emmanuel~N.
  Saridakis.
\newblock {Viscous Cosmology for Early- and Late-Time Universe}.
\newblock {\em Int. J. Mod. Phys.}, D26:1730024, 2017.

\bibitem{Eckart1940a}
Carl Eckart.
\newblock {The Thermodynamics of irreversible processes. 3.. Relativistic
  theory of the simple fluid}.
\newblock {\em Phys. Rev.}, 58:919--924, 1940.

\bibitem{Li2009}
Baojiu Li and John~D. Barrow.
\newblock {Does Bulk Viscosity Create a Viable Unified Dark Matter Model?}
\newblock {\em Phys. Rev.}, D79:103521, 2009.

\bibitem{Avelino2009}
Arturo Avelino and Ulises Nucamendi.
\newblock {Can a matter-dominated model with constant bulk viscosity drive the
  accelerated expansion of the universe?}
\newblock {\em JCAP}, 0904:006, 2009.

\bibitem{Velten2011}
Hermano Velten and Dominik~J. Schwarz.
\newblock {Constraints on dissipative unified dark matter}.
\newblock {\em JCAP}, 1109:016, 2011.

\bibitem{VeltenSchwarz2012}
Hermano Velten and Dominik Schwarz.
\newblock {Dissipation of dark matter}.
\newblock {\em Phys. Rev.}, D86:083501, 2012.

\bibitem{VeltenSchwarzFabrisEtAl2013}
H.~Velten, D.~J. Schwarz, J.~C. Fabris, and W.~Zimdahl.
\newblock {Viscous dark matter growth in (neo-)Newtonian cosmology}.
\newblock {\em Phys. Rev.}, D88(10):103522, 2013.

\bibitem{Velten2013}
Hermano Velten, Jiaxin Wang, and Xinhe Meng.
\newblock {Phantom dark energy as an effect of bulk viscosity}.
\newblock {\em Phys. Rev.}, D88:123504, 2013.

\bibitem{AvelinoLeyvaUrena-Lopez2013}
Arturo Avelino, Yoelsy Leyva, and L.~Arturo Urena-Lopez.
\newblock {Interacting viscous dark fluids}.
\newblock {\em Phys. Rev.}, D88:123004, 2013.

\bibitem{CruzLepeLeyvaEtAl2014}
Norman Cruz, Samuel Lepe, Yoelsy Leyva, Francisco Peña, and Joel Saavedra.
\newblock {No stable dissipative phantom scenario in the framework of a
  complete cosmological dynamics}.
\newblock {\em Phys. Rev.}, D90(8):083524, 2014.

\bibitem{Leyva2017}
Yoelsy Leyva and Mirko Sepúlveda.
\newblock {Bulk viscosity, interaction and the viability of phantom solutions}.
\newblock {\em Eur. Phys. J.}, C77(6):426, 2017.

\bibitem{Israel1979}
W.~Israel and J.~M. Stewart.
\newblock {Transient relativistic thermodynamics and kinetic theory}.
\newblock {\em Annals Phys.}, 118:341--372, 1979.

\bibitem{Maartens1996}
Roy Maartens.
\newblock {Causal thermodynamics in relativity}.
\newblock 1996.

\bibitem{Zakari1993}
Mohamed Zakari and David Jou.
\newblock {Equations of state and transport equations in viscous cosmological
  models}.
\newblock {\em Phys. Rev.}, D48(4):1597, 1993.

\bibitem{Oliveira1988}
H.~P. de~Oliveira and J.~M. Salim.
\newblock {Nonequilibrium Friedmann Cosmologies}.
\newblock {\em Acta Phys. Polon.}, B19:649, 1988.

\bibitem{Pavon1991}
D.~Pavon, J.~Bafaluy, and D.~Jou.
\newblock {Causal Friedmann-Robertson-Walker cosmology}.
\newblock {\em Class. Quant. Grav.}, 8:347--360, 1991.

\bibitem{Cruz2017b}
Norman Cruz, Samuel Lepe, and Francisco Peña.
\newblock {Crossing the phantom divide with dissipative normal matter in the
  Israel–Stewart formalism}.
\newblock {\em Phys. Lett.}, B767:103--109, 2017.

\bibitem{Chapman1991}
S.~Chapman and T.~G. Cowling.
\newblock {\em The Mathematical Theory of Non-uniform Gases}.
\newblock January 1991.

\bibitem{Hiscock1991}
W.~A. Hiscock and J.~Salmonson.
\newblock {Dissipative Boltzmann-Robertson-Walker cosmologies}.
\newblock {\em Phys. Rev.}, D43:3249--3258, 1991.

\bibitem{Jeon1995}
Sangyong Jeon.
\newblock {Hydrodynamic transport coefficients in relativistic scalar field
  theory}.
\newblock {\em Phys. Rev.}, D52:3591--3642, 1995.

\bibitem{Jeon1996}
Sangyong Jeon and Laurence~G. Yaffe.
\newblock {From quantum field theory to hydrodynamics: Transport coefficients
  and effective kinetic theory}.
\newblock {\em Phys. Rev.}, D53:5799--5809, 1996.

\bibitem{Bastero-Gil2012}
Mar Bastero-Gil, Arjun Berera, Rafael Cerezo, Rudnei~O. Ramos, and Gustavo~S.
  Vicente.
\newblock {Stability analysis for the background equations for inflation with
  dissipation and in a viscous radiation bath}.
\newblock {\em JCAP}, 1211:042, 2012.

\bibitem{Ren2006}
Jie Ren and Xin-He Meng.
\newblock {Cosmological model with viscosity media (dark fluid) described by an
  effective equation of state}.
\newblock {\em Phys. Lett.}, B633:1--8, 2006.

\bibitem{Hu2006}
Ming-Guang Hu and Xin-He Meng.
\newblock {Bulk viscous cosmology: statefinder and entropy}.
\newblock {\em Phys. Lett.}, B635:186--194, 2006.

\bibitem{Kremer2012}
Gilberto~M. Kremer and Octavio A.~S. Sobreiro.
\newblock {Bulk viscous cosmological model with interacting dark fluids}.
\newblock {\em Braz. J. Phys.}, 42:77--83, 2012.

\bibitem{Acquaviva2014}
G.~Acquaviva and A.~Beesham.
\newblock {Nonlinear bulk viscosity and the stability of accelerated expansion
  in FRW spacetime}.
\newblock {\em Phys. Rev.}, D90(2):023503, 2014.

\bibitem{Haro2015}
Jaume Haro and Supriya Pan.
\newblock {Bulk viscous quintessential inflation}.
\newblock 2015.

\bibitem{Wang2017}
Deng Wang, Yang-Jie Yan, and Xin-He Meng.
\newblock {Constraining viscous dark energy models with the latest cosmological
  data}.
\newblock {\em Eur. Phys. J.}, C77(10):660, 2017.

\bibitem{Maartens1997}
Roy Maartens and Vicenc Mendez.
\newblock {Nonlinear bulk viscosity and inflation}.
\newblock {\em Phys. Rev.}, D55:1937--1942, 1997.

\bibitem{LeonLeyvaSocorro2014}
Genly Leon, Yoelsy Leyva, and J.~Socorro.
\newblock {Quintom phase-space: beyond the exponential potential}.
\newblock {\em Phys. Lett.}, B732:285--297, 2014.

\bibitem{Zimdahl2000}
Winfried Zimdahl and Diego Pavon.
\newblock {Expanding universe with positive bulk viscous pressures?}
\newblock {\em Phys. Rev.}, D61:108301, 2000.

\bibitem{Misner1973}
Charles~W. Misner, K.~S. Thorne, and J.~A. Wheeler.
\newblock {\em {Gravitation}}.
\newblock W. H. Freeman, San Francisco, 1973.

\bibitem{ThomasKoppSkordis2016}
Daniel~B. Thomas, Michael Kopp, and Constantinos Skordis.
\newblock {Constraining dark matter properties with Cosmic Microwave Background
  observations}.
\newblock {\em Astrophys. J.}, 830(2):155, 2016.

\bibitem{Lepe2017}
Samuel Lepe, Giovanni Otalora, and Joel Saavedra.
\newblock {Dynamics of viscous cosmologies in the full Israel-Stewart
  formalism}.
\newblock {\em Phys. Rev.}, D96(2):023536, 2017.

\end{thebibliography}
\end{document}